\documentclass[sigconf,natbib=true]{acmart}
\usepackage[ruled,linesnumbered]{algorithm2e}
\usepackage{multirow}
\usepackage{makecell}
\usepackage{enumitem}
\setitemize[1]{noitemsep,partopsep=0pt,parsep=0pt,topsep=0pt, leftmargin=10pt,
rightmargin=0pt}
\AtBeginDocument{%
  \providecommand\BibTeX{{%
    \normalfont B\kern-0.5em{\scshape i\kern-0.25em b}\kern-0.8em\TeX}}}

\setcopyright{acmcopyright}
\copyrightyear{2018}
\acmYear{2018}
\acmDOI{10.1145/1122445.1122456}

\acmConference[Woodstock '18]{Woodstock '18: ACM Symposium on Neural
  Gaze Detection}{June 03--05, 2018}{Woodstock, NY}
\acmBooktitle{Woodstock '18: ACM Symposium on Neural Gaze Detection,
  June 03--05, 2018, Woodstock, NY}
\acmPrice{15.00}
\acmISBN{978-1-4503-XXXX-X/18/06}

\begin{document}

\title{Global-Distribution Aware Scenario-Specific Variational Representation Learning Framework}


\author{Moyu Zhang}
\affiliation{%
  \institution{Alibaba Group}
  \city{Beijing}
  \state{Beijing}
  \country{China}
}
\email{zhangmoyu@bupt.cn}

\author{Yujun Jin}
\affiliation{%
  \institution{Alibaba Group}
  \city{Beijing}
  \state{Beijing}
  \country{China}
}
\email{jinyujun.jyj@alibaba-inc.com}

\author{Jinxin Hu}
\authornote{Corresponding Author}
\affiliation{%
  \institution{Alibaba Group}
  \city{Beijing}
  \state{Beijing}
  \country{China}
}
\email{jinxin.hjx@alibaba-inc.com}

\author{Yu Zhang}
\affiliation{%
  \institution{Alibaba Group}
  \city{Beijing}
  \state{Beijing}
  \country{China}
}
\email{daoji@lazada.com}

\begin{abstract}
With the emergence of e-commerce, the recommendations provided by commercial platforms must adapt to diverse scenarios to accommodate users' varying shopping preferences. Current methods typically use a unified framework to offer personalized recommendations for different scenarios. However, they often employ shared bottom representations, which partially hinders the model's capacity to capture scenario uniqueness. Ideally, users and items should exhibit specific characteristics in different scenarios, prompting the need to learn scenario-specific representations to differentiate scenarios. Yet, variations in user and item interactions across scenarios lead to data sparsity issues, impeding the acquisition of scenario-specific representations. To learn robust scenario-specific representations, we introduce a \textbf{G}lobal-Distribution Aware \textbf{S}cenario-Specific \textbf{V}ariational \textbf{R}epresentation Learning Framework (GSVR) that can be directly applied to existing multi-scenario methods. Specifically, considering the uncertainty stemming from limited samples, our approach employs a probabilistic model to generate scenario-specific distributions for each user and item in each scenario, estimated through variational inference (VI). Additionally, we introduce the global knowledge-aware multinomial distributions as prior knowledge to regulate the learning of the posterior user and item distributions, ensuring similarities among distributions for users with akin interests and items with similar side information. This mitigates the risk of users or items with fewer records being overwhelmed in sparse scenarios. Extensive experimental results affirm the efficacy of GSVR in assisting existing multi-scenario recommendation methods in learning more robust representations.
\end{abstract}


\maketitle

\section{Introduction}
In recent years, the mainstream adoption of e-shopping has led to the emergence of numerous e-commerce platforms such as Amazon, Taobao, TikTok, and others \cite{back1, back2, back3, back4, causalint, adl}. To accommodate increasingly diverse user shopping preferences, modern commercial platforms are required to deliver personalized services for a variety of shopping environments. In practice, significant user and item overlap exists across scenarios in most current e-commerce platforms. Thus, despite performance differences, notable correlations persist among users and items in different scenarios. Mainstream multi-scenario approaches typically employ a unified recommendation modeling framework, encompassing two categories: 

1) Scenario-specific network structures, influenced by multi-task learning (MTL) \cite{mtl1, mtl2, mtl3}. Each scenario is treated as a distinct task, with a common network capturing inter-scenario correlations and separate networks modeling the unique aspects of each scenario. 2) Scenario-adaptive parameter network structures \cite{pepnet, adasparse}, which differ from MTL by acknowledging feature space variations. These methods apply scenario data directly to the core embedding and prediction layers, enabling dynamic adjustments of the feature space and prediction strategy in response to scenario shifts.

However, current multi-scenario modeling structures commonly utilize shared underlying representations, relying solely on predefined scenario IDs to differentiate scenarios. This approach hinders the accurate distinction of scenario differences by the model. Hence, it is intuitive to learn scenario-specific representations for users and items to enhance the model's scenario discrimination capability. Nonetheless, real-world scenarios often exhibit data sparsity issues, which results in greater uncertainty of users and items. Consequently, scenario-specific representations suffer from inadequate learning, leading to negative transfer problems and weakening the model's ability to capture changes in user interests within sparse scenarios. Effectively modeling the uncertainty of users and items in each scenario remains a challenge in multi-scenario modeling.

To tackle the above challenge, we introduce a probabilistic model that learns the distribution of each user and item in different scenarios to represent their uncertainty. We calculate the expected value of the distribution as the final prediction score, more effectively reducing uncertainty. Furthermore, recognizing that scenario-specific representations are impacted by data sparsity, we propose leveraging global-knowledge to aid scenario-specific distribution modeling. Specifically, we propose to utilize the multi-experts network \cite{ple} to model user sequences, with shared experts representing global behavior modes and scenario-specific experts representing scenario-specific behavior modes. Based on extracted behavior modes, we design a behavior-mode aware multinomial distribution as a prior during Bayesian inference, constraining the model’s learning of posterior user distributions. Meanwhile, to better share the global and statistical strength among items, we propose to parameterize priors with neural network and the attributes of items as inputs to constraining the model’s learning of posterior item distributions. To further avoid overfitting, we also use the standard normal distribution as a kind of prior knowledge to constrain model training.

The contributions of our paper can be summarized as follows:
\begin{itemize}
\item To the best of our knowledge, we are the first study to learn robust scenario-specific representations by building a probabilistic model that generates a distribution for each user and item to track differences across various scenarios.
\item We design the global-knowledge aware prior multinomial distributions to constrain the model to generate similar posterior distributions for users with similar behavior modes and items with similar side information.
\item GSVR is a general framework that can be incorporated into existing methods. Our experimental results demonstrate the superiority and compatibility of the GSVR framework.
\end{itemize}  

\section{Preliminaries}
Given a set of scenarios $\mathcal{S}=\left\{s\right\}_{m=1}^{N_s}$ with a shared feature space $\boldsymbol{x}$ and label space $\mathcal{Y}$, the multi-scenario recommendation task aims to devise a unified ranking formula $\mathcal{F}: \boldsymbol{x} \rightarrow  \mathcal{Y}$, to concurrently provide accurate, personalized recommendations across $M$ scenarios. Each instance contains the features $\boldsymbol{x}$ implying the information of $\left\{u, v, b_u, c_v, s \right\}$, where $u$ denotes the user ID index, $v$ denotes the item ID index, and ${s}$ denotes the scenario ID index. $b_u = \left\{b_i\right\}_{i=0}^{N_b}$ denote the user's historical behavior sequence and $N_b$ represents the number of the user's historical behaviors. $c$ denotes the side features of the item $v$. Within this paper, the user $u$ and item $v$ can be embed as a scenario-shared embedding vector and a scenario-specific embedding vector, respectively. The shared embedding of user and item can be expressed as $\boldsymbol{u}$ and $\boldsymbol{v}$, and the scenario-specific embedding of user and item can be denoted as $\boldsymbol{u}_{s}$ and $\boldsymbol{v}_{s}$ for the specific scenario ${s}$. Similarly, $c$ can be embed as $\boldsymbol{c}$ and $\boldsymbol{c}_{s}$. Since user sequences are generally collected from all domains, we use shared embeddings to represent the clicked items in the sequence as $\boldsymbol{b}_u = \left\{\boldsymbol{b}_i\right\}_{i=0}^{N_b}$. Mathematically, the task can be illustrated below:
\begin{gather}
\hat{y} = P(y=1| \boldsymbol{x}) =  \mathcal{F}(\boldsymbol{u}, \boldsymbol{u_{s}}, \boldsymbol{v}, \boldsymbol{v_{s}}, \boldsymbol{b}_u, \boldsymbol{c}, \boldsymbol{c}_{s},  \boldsymbol{s})
\end{gather} 

\section{Method}
To help model the uncertainty of the users and items in sparse scenarios, we build a probabilistic representation learning framework to generate a distribution for each user and item in the corresponding scenario. Simultaneously, we use the global-distribution extracted from all scenarios as a prior distribution for the embedding parameter space, which constrains the distribution estimate. 
\subsection{GSVR: Distribution Estimate Framework} 

Current multi-scenario recommendation methods utilize the shared representations to forecast whether uses will click the target item in different scenarios. Intuitively, users and items should possess specific traits in different scenarios, and we hope to learn a set of scenario-specific representations for each user and item at the bottom level, increase the information differences between scenarios from the bottom up, and thus enhance the ability to model the differences between scenarios. However, although there is overlap between users and items in many scenarios, the data in each scenario is relatively sparse, which makes it difficult for us to fully learn the scenario-specific representations. To alleviate the problem of data sparsity, we believe that distribution estimation should be used instead of the original point estimation to model the uncertainty in scenarios, and estimation of the posterior distribution over the latent space $\boldsymbol{z}$, denoted as $p(\boldsymbol{z}|\boldsymbol{x})$, is necessary. $\boldsymbol{z}$ contains the scenario-shared and scenario-specific embedding parameters. Then the variational inference is applied to reformulate the target of the recommendation task as $p_{\phi, \theta}(y=1|\boldsymbol{x}, \boldsymbol{z})$, where $\phi$ denotes the parameters of the embedding vectors and $\theta$ denotes the parameters of neural networks. Due to the intractability of the true posterior distribution, a recognition model $q_{\phi}(\boldsymbol{z}|\boldsymbol{x})$ can be utilized to approximate the true distribution, where $q_{\phi}(\boldsymbol{z}|\boldsymbol{x})$ can be interpreted as a probabilistic encoder generating a distribution (such as a Gaussian distribution) for each feature. Based on Bayes theorem, the posterior of $\boldsymbol{z}$ can be represented as $p(\boldsymbol{z}|\boldsymbol{x}) = \frac{p(\boldsymbol{z}, \boldsymbol{x})}{p(\boldsymbol{x})}$. Thus, based on Jensen's inequality, the evidence lower bound (ELBO) of the marginal likelihood $p(\boldsymbol{x})$ can be obtained as follows:
\begin{gather}
\begin{aligned}
log \, p(\boldsymbol{x}) &= log \int p(\boldsymbol{x}, \boldsymbol{z}) d \boldsymbol{z}  = log \int p(\boldsymbol{x}, \boldsymbol{z}) \frac{q_{\phi}(\boldsymbol{z}| \boldsymbol{x})}{q_{\phi}(\boldsymbol{z}| \boldsymbol{x})} d \boldsymbol{z} \\
& \ge \mathbb{E}_{q_{\phi}(\boldsymbol{z} | \boldsymbol{x})} [log p(\boldsymbol{x}, \boldsymbol{z}) - log q_{\phi}(\boldsymbol{z} | \boldsymbol{x})] \\
&= \mathbb{E}_{q_{\phi}(\boldsymbol{z} | \boldsymbol{x})} [log p_{\theta}(\boldsymbol{x}| \boldsymbol{z})] - KL(q_{\phi}(\boldsymbol{z} | \boldsymbol{x}) || p(\boldsymbol{z}))
\end{aligned}
\end{gather}
where the first term can be regarded as an expected negative reconstruction error (denoted as $\mathcal{L}_{pre}$), while the second term serves as a regularizer to constrain the approximate posterior distribution $q_{\phi}(\boldsymbol{z} | \boldsymbol{x})$ via the prior distribution $p(\boldsymbol{z})$. The objective function $\mathcal{L}(\phi, \theta)$ is naturally equivalent to the ELBO, and maximizing $p(\boldsymbol{x})$ is equivalent to minimizing the lower bound. 

With the help of the Mean-field Theory \cite{mean}, we assume features in $\boldsymbol{x}$ are mutually independent and each feature is governed by distinct factors in the variational density. Consequently, the objective function of our model can be reformulated as follows:
\begin{gather}
\begin{aligned}
\mathcal{L}(\phi, \theta)  = \mathcal{L}_{pre} &- \alpha [KL(q_{{\phi}_{u}}(\boldsymbol{z}_{u_{s}} | {u_{s}}) || p(\boldsymbol{z}_{u_{s}})) + \\
& KL(q_{{\phi}_{v}}(\boldsymbol{z}_{v_{s}} | {v_{s}}) || p(\boldsymbol{z}_{v_{s}}))]
\end{aligned}
\end{gather}
where $p(\boldsymbol{z}_{u_{s}})$ and $p(\boldsymbol{z}_{v_{s}})$ usually denote the normal Gaussian distribution.  $\alpha$ allows to achieve a better balance between the latent space independence and reconstruction errors to achieve better prediction performance, as shown in ${\beta}$-VAE \cite{b_vae}. However, we suspect that the fixed prior distribution limits the generalization ability of the model due to the significant variation between different users and different items. As a solution, we propose parameterizing $p(\boldsymbol{z}_{v_{s}})$ of the item $v$ as $p_{{\phi}_{c}}(\boldsymbol{z}_{v_{s}}|{\boldsymbol{c}_{s}})$, where ${c_{s}}$ denotes the side features related to the item $v$ in the scenario $s$. In this way, items with similar side features will have similar prior distributions restrictions. As for users, since they do not have relatively strong personalized attached features like brands or categories, their characteristics are often reflected in behavior sequences, so we hope to model users with similar sequences as having similar distributions. Considering that users have different interests reflected by sequences in different scenarios, we use the MoE structure to extract behavior patterns from user behavior sequences, i.e., $p(\boldsymbol{z}_{u_{s}}) \rightarrow p_{{\phi}_{b}}(\boldsymbol{z}_{u_{s}}|\left\{b_i\right\}_{i=0}^{N_b})$. The specific process is as follows:
\begin{gather}
\boldsymbol{h} = f (\left\{b_i\right\}_{i=0}^{N_b}) \\
\boldsymbol{\alpha}_c = G_c(\boldsymbol{h}), \quad \boldsymbol{\alpha}_s = G_s(\boldsymbol{h} \oplus \boldsymbol{s}) \\
\boldsymbol{b}_{u_s} = (\sum_{1 \leq i < d_c} \boldsymbol{\alpha}_{c_i} E_{c_i}(\boldsymbol{h})) \oplus (\sum_{1 \leq i < d_s} \boldsymbol{\alpha}_{s_i} E_{s_i}(\boldsymbol{h}  \oplus \boldsymbol{s} ))
\end{gather}
where $\boldsymbol{b}_{u_s}$ denotes the representation of extracted behavior modes. $\boldsymbol{\alpha}_c$ and $\boldsymbol{\alpha}_s$ denote the weight of each shared and specific expert, respectively. $d_c$ and $d_s$ denotes the number of shared and specific experts, respectively. The function can be reformulated as follows:
\begin{gather}
\begin{aligned}
\mathcal{L}(\phi, \theta)  = \mathcal{L}_{pre} &- \alpha[KL(q_{{\phi}_{u}}(\boldsymbol{z}_{u_{s}} | {u_{s}}) || p_{{\phi}_{b}}(\boldsymbol{z}_{u_{s}}|\boldsymbol{e}_s)+ \\
&KL(q_{{\phi}_{v}}(\boldsymbol{z}_{v_{s}} | {v_{s}}) || p_{{\phi}_{c}}(\boldsymbol{z}_{v_{s}}|c_s))]
\end{aligned}
\end{gather}

\textbf{\emph{Re-parameterize Trick.}} Next, we apply a re-parameterize trick to generate the posterior distributions as below:
\begin{gather}
q_{{\phi}_{u}}(\boldsymbol{z}_{u_s} | {u_s}) = \mathcal{N} (\boldsymbol{\mu}_{u_s}, \boldsymbol{\sigma}^2_{u_s}) \\
q_{{\phi}_{v}}(\boldsymbol{z}_{v_s} | {v_s}) = \mathcal{N} (\boldsymbol{\mu}_{v_s}, \boldsymbol{\sigma}^2_{v_s})
\end{gather}
where $\mu_{u_s}$ or $\mu_{v_s}$, and $\sigma_{u_s}$ or $\sigma_{v_s}$, are obtained from scenario-specific representations of user ID or item ID with DNNs. Similarly, we can obtain the parameterized prior distributions as follows:
\begin{gather}
p_{{\phi}_{c}}(\boldsymbol{z}_{v_{s}}|{c_{s}}) = \mathcal{N} (\boldsymbol{\mu}_{c_{s}},  \boldsymbol{\sigma}^2_{c_{s}}) \\
p_{{\phi}_{b}}(\boldsymbol{z}_{u_{s}}|\left\{b_i\right\}_{i=0}^{N_b}) = \mathcal{N} (\boldsymbol{\mu}_b, \boldsymbol{\sigma}^2_b)
\end{gather}
where $\mu_{c_{s}}$ and $\sigma_{c_{s}}$ are obtained from the representations of side features related to the target item with DNNs. $\mu_b$ and $\sigma^2_b$ are obtained from the representation of extracted behavior mode $\boldsymbol{e}_s$ with DNNs. 

\subsection{Training Phase}
\subsubsection{Prediction} 
Based on the above estimated distributions, we can generate scenario-specific embedding vectors for users and items by random sampling, as follows:
\begin{gather}
\boldsymbol{u_{s}} = \boldsymbol{\mu}_{u_s} +  \boldsymbol{\sigma}_{u_s} \odot \boldsymbol{\epsilon}_{u_s}, \quad \boldsymbol{\epsilon}_{u_s} \in \mathcal{N} (0,  \boldsymbol{I}) \\
\boldsymbol{v_{s}} = \boldsymbol{\mu}_{v_s} +  \boldsymbol{\sigma}_{v_s} \odot \boldsymbol{\epsilon}_{v_s}, \quad \boldsymbol{\epsilon}_{v_s} \in \mathcal{N} (0,  \boldsymbol{I})
\end{gather}
Then, we conduct predictions as follows:
\begin{gather}
\hat{y}  = f_{\theta}(\boldsymbol{u}, \boldsymbol{u_{s}}, \boldsymbol{v},  \boldsymbol{v_{s}}, \boldsymbol{b}_{u_s}, \boldsymbol{c}, \boldsymbol{c}_s,  \boldsymbol{s})
\end{gather}
where $\hat{y}$ is the prediction probability value. $f_{\theta}(\cdot)$ represents the prediction layer, which can be DNNs and other structures. The resulting reconstruction error can be calculated by:
\begin{equation}
\mathcal{L}_{pre} = -\begin{matrix} \frac{1}{L} \sum_{1 \leq i \leq L} y log \hat{y}_i + (1-y)log(1-\hat{y}_i) \end{matrix}
\end{equation}
where $\hat{y}_i$ denotes the predicted probability of sampled $\epsilon_u$ and $\epsilon_v$. $L$ is the Monte Carlo sampling number for each record. To avoid the over-fitting issue, we add regularization terms on both distributions of users and items by forcing the parameterized distributions to be close to a standard normal Gaussian distributions. In this way, the final objective function can be reformulated as below:
\begin{gather}
\begin{aligned}
& \mathcal{L}(\phi, \theta)  = \mathcal{L}_{re}  \\
& - \alpha[KL(q_{{\phi}_{u}}(\boldsymbol{z}_{u_{s}} | {u_{s}}) || p_{{\phi}_{b}}(\boldsymbol{z}_{u_{s}}|\boldsymbol{e}_s)+ KL(q_{{\phi}_{v}}(\boldsymbol{z}_ | {v_{s}}) || p_{{\phi}_{c}}(\boldsymbol{z}_{v_{s}}|c_s))] \\
&- \alpha [KL(p_{{\phi}_{b}}(\boldsymbol{z}_{u_{s}}|\boldsymbol{e}_s) || \mathcal{N} (0,  \boldsymbol{I}))  + KL(p_{{\phi}_{c}}(\boldsymbol{z}_{v_{s}}|c_s)) || \mathcal{N} (0,  \boldsymbol{I}))] 
\end{aligned}
\end{gather}

\subsection{Inferring Phase}
After training phase, we can get the distribution of each user and item, and regard the mean values of the multinomial distribution as the final representation to conduct the prediction $\hat{y}$ as follows:
\begin{gather}
\boldsymbol{u_{s}} = \boldsymbol{\mu}_{u_s}, \quad \boldsymbol{v_{s}} = \boldsymbol{\mu}_{v_s} \\
\hat{y}  = f_{\theta}(\boldsymbol{u}, \boldsymbol{u_{s}}, \boldsymbol{v}, \boldsymbol{v_{s}},  \boldsymbol{b}_{u_s}, \boldsymbol{c}, \boldsymbol{c}_s,  \boldsymbol{s})
\end{gather}
where $\boldsymbol{u_{s}}$ and $\boldsymbol{v_{s}}$ are obatined by replacing the random sampling vectors $\boldsymbol{\epsilon}_{u_s}$ and $\boldsymbol{\epsilon}_{v_s}$ with $\boldsymbol{0}$.  

\begin{table}[t]
	\small
	\centering
	\begin{tabular}{ccccccccccccc}
    		\hline Scenarios & \#A1 & \#A2 & \#A3 & \#A4\\
		\hline \#User & 10,430,696 & 4,877,528& 3,703,052 & 1,067,512\\
		\#Item & 1,972,477 & 1,535,663 & 1,289,892 & 1,330,079 \\
		\#Click & 15,870,365 & 9,220,309 & 1,966,885 & 7,521,706\\
		\#Impression & 860M & 265M & 76M & 56M \\
		\hline
	\end{tabular}
	\caption{Statistics of the industrial dataset.}
	\label{industrial_dataset}
\end{table} 

\begin{table}[t]
	\small
	\centering
	\begin{tabular}{cccc}
    		\hline Scenarios & \#B1 & \#B2 & \#B3  \\
		\hline \#User & 91,488 & 2,612 & 154,024   \\
		\#Item & 535,711 & 198,651 &537,937  \\
		\#Click & 1,291,063 & 28,022 &1,998,618 \\
		\#Impression & 32,236,951 & 639,897 &52,439,671 \\
		\hline
	\end{tabular}
	\caption{Statistics of the Ali-CCP dataset.}
	\label{public_dataset}
\end{table}

\section{Experiments}
\subsection{Experimental Setup}
\subsubsection{Datasets}
The descriptions and statistics of two dataset are detailed in Table \ref{industrial_dataset} and \ref{public_dataset}, respectively. \textbf{$\bullet$ Industrial Dataset}. We gathered a dataset from an international advertising platform that spans multiple scenarios from April 26, 2025, to May 16, 2025. This dataset encompasses 4 scenarios, denoted as \#A1 through \#A4. \textbf{$\bullet$ Ali-CCP  \footnote{https://tianchi.aliyun.com/dataset/408}}.  It is widely used for multi-scenario recommendations \cite{maria, plate, sass}, which was collected Taobao’s recommender system under three scenarios, denoted as \#B1, \#B2, and \#B3 for simplicity.

\subsection{Competitors}
\textbf{$\bullet$ SharedBottom}.  An individual DNN is utilized for each scenario. \\
\textbf{$\bullet$ PLE} \cite{ple}.  It organizes experts into scenario-specific and scenario-shared groups for avoiding seesaw phenomenon. \\
\textbf{$\bullet$ AESM$^2$}  \cite{aesm2}.  It proposes a novel MMoE-based model with automatic search towards the optimal network structure.  \\
\textbf{$\bullet$ AdaSparse} \cite{adasparse}.  It learns adaptively sparse structures for multi-scenario prediction and prunes redundant neurons. \\
\textbf{$\bullet$ PEPNet} \cite{pepnet}.  It dynamically scales the bottom-layer embeddings and the top-layer DNN hidden units. \\
\textbf{$\bullet$ MARIA} \cite{maria}. It designs feature scaling, refinement, and correlation modeling, to enable discriminative feature learning. 

\begin{table}[t]
	\caption{Prediction performance on the industrial dataset. * indicates p-value < 0.05 in the significance test.}
	\begin{tabular}{c|c|c|c|c}
    \toprule
    \multirow{3}{*}{Method}&
    \multicolumn{4}{c}{The industrial dataset (S-GAUC)}\cr
    \cmidrule(lr){2-5}
    &\#A1&\#A2&\#A3&\#A4\cr
    \midrule
    SharedBottom  & 0.6282 & 0.6756 & 0.6313 & 0.5824 \cr
    PLE   & 0.6356 & 0.6829 & 0.6398    & 0.5916 \cr
    AESM$^2$  &0.6368  &0.6844  & 0.6413  &0.5931 \cr
    AdaSparse   & 0.6389 & 0.6871 & 0.6441  & 0.5952  \cr
    PEPNet & 0.6403 & 0.6883 & 0.6452 & 0.5991 \cr 
    MARIA  & 0.6412 & 0.6901 & 0.6471  & 0.6002 \cr
    \midrule
    PEPNet+GSVR& \textbf{0.6424}* & \textbf{0.6912}* & \textbf{0.6489}* &  \textbf{0.6027}*\cr
    MARIA+GSVR& \textbf{0.6459}* & \textbf{0.6937}* & \textbf{0.6526}* &  \textbf{0.6068}*\cr
    \bottomrule
    \end{tabular}
	\label{result_ind}
\end{table} 

\begin{table}[t]
	\caption{Prediction performance on the Ali-CCP dataset. * indicates p-value < 0.05 in the significance test.}
	\begin{tabular}{c|c|c|c}
    \toprule
    \multirow{3}{*}{Method}&
    \multicolumn{3}{c}{Ali-CCP dataset (AUC)}\cr
    \cmidrule(lr){2-4}
    &\#B1&\#B2&\#B3\cr
    \midrule
    SharedBottom  & 0.5685 & 0.5881 & 0.5817 \cr
    PLE   & 0.5786 & 0.5828 & 0.5893   \cr
    AESM$^2$   & 0.5798 & 0.5963 & 0.5927   \cr
    AdaSparse   & 0.5808 & 0.5986 & 0.6024\cr
    PEPNet & 0.5839 & 0.6101 & 0.6052  \cr 
    MARIA  & 0.5864 & 0.6108 & 0.6068  \cr
    \midrule
    PEPNet+GSVR & \textbf{0.5921}* & \textbf{0.6149}* & \textbf{0.6115}*  \cr
    MARIA+GSVR & \textbf{0.5937}* & \textbf{0.6163}* & \textbf{0.6127}*  \cr
    \bottomrule
    \end{tabular}
	\label{result_pub}
\end{table} 

\textbf{Implementation Details}. We implement all the models based on the TensorFlow framework \cite{tensorflow}. We use Adam \cite{adam} for optimization with an initial learning rate of 0.001 and a decay rate of 0.9. The batch size is set as 512 and the embedding size is fixed to 40. Xavier initialization \cite{xavier} is used. All methods use a three-layer neural network with hidden sizes of [256, 128, 64].

\textbf{Evaluation Metric}. We employ the area under the ROC curve (AUC) as the metric for the Ali-CCP dataset. For our industrial dataset, we adopt the session-weighted AUC (S-GAUC) \cite{sfpnet}. Mann-Whitney U test \cite{m-test} is conducted under AUC and S-GAUC metrics. 

\subsection{Effectiveness Verification}
This section compares GSVR with baselines using different backbone models. Tables \ref{result_ind} and \ref{result_pub} display the average results, highlighting the highest score in bold. It is evident that while PEPNet and MARIA may achieve the best performance in the baseline, GSVR significantly enhances the accuracy of these methods in predicting each scenario on the underlying superposition. PEPNet and MARIA aim to enhance the contribution of scenario features to final prediction through optimizing models' structures, overlooking the limited information from the underlying scenario features themselves.
 
\subsection{Ablation Study}
To get deep insights into contributions of components, we conduct ablation studies by applying multiple
variants to MARIA: \textbf{ GSVR(Distinct)} directly learns distinct representations for each scenario. \textbf{GSVR(Uniform)} uses the uniform prior distributions as VAE does. \textbf{GSVR(R-MoE)}.  We removed the expert networks and directly input the representations of sequences for re-parametrizing.

Tables \ref{result_abla} reports the results of three variants. GSVR outperforms GSVR(R-MoE) as the expert network extracts behavior patterns from user sequences to aid in user representation learning. GSVR(Uniform) demonstrates the better performance than baselines, yet it is still inferior to GSVR, emphasizing the advantage of utilizing a global-distribution to facilitate learning single-scenario distribution representations, thus mitigating the overfitting phenomenon caused by data sparsity during small-scenario personalization. Lastly, GSVR(Distinct) performs the worst among all variants, strongly indicating the difficulty for the model to directly learn a set of representation vectors for each scenario.

\subsection{Hyper-Parameters Sensitivity Analysis}
\begin{figure}[t]
  \centering
  \includegraphics[width=\linewidth]{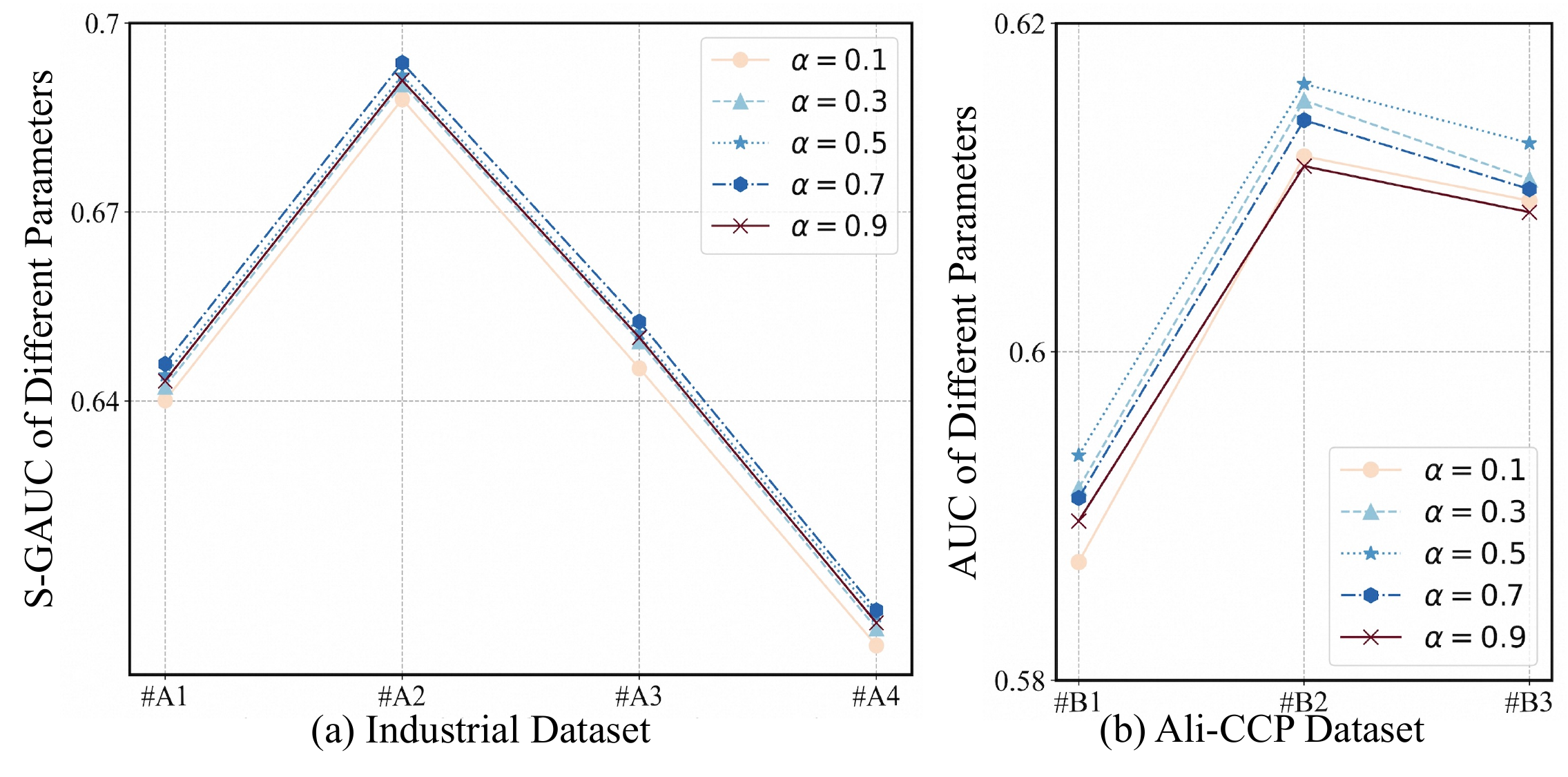}
  \caption{The performance of MARIA+GSVR under different predefined hyper-parameters on the both datasets.}
  \label{hyper}
\end{figure}

We evaluate GSVR’s performance with five different values: 0.1, 0.3, 0.5, 0.7, and 0.9. The experimental results on the two datasets are shown in Figure \ref{hyper}. The main task is intended to predict users' responses to target items, not to maximize the probability likelihood of the simulated distribution. Therefore, the closer the value of $\alpha$ is to 1, the more likely our task will deviate from the original intention, leading to decreased prediction performance of the model. On the other hand, too small a value may cause our regularization to lose its effect and degenerate into an ordinary deep model, leading to overfitting. Therefore, setting the value of $\alpha$ in the range of 0.5-0.7 can achieve a more universal effect.

\begin{table}[t]
	\caption{Performance of variants on the industrial dataset. }
	\begin{tabular}{c|c|c|c|c}
    \toprule
    \multirow{3}{*}{Method}&
    \multicolumn{4}{c}{The industrial dataset (S-GAUC)}\cr
    \cmidrule(lr){2-5}
    &\#A1&\#A2&\#A3&\#A4\cr
    \midrule
    GSVR&  \textbf{0.6459}* & \textbf{0.6937}* & \textbf{0.6526}* &  \textbf{0.6068}*\cr
    GSVR(Distinct)   & 0.6416 & 0.6899 & 0.6473    & 0.5997 \cr
    GSVR(Uniform)  &0.6430  &0.6914  & 0.6499  &0.6009 \cr
    GSVR(R-MoE)   & 0.6442 & 0.6929 & 0.6513  & 0.6041  \cr
    \bottomrule
    \end{tabular}
	\label{result_abla}
\end{table}

\subsection{Online A/B Testing Results}
We conducted an online A/B test on an online e-commerce platform. The base model is PEPNet. We averaged the results of the 4 scenarios from April 21 to 30, 2025, to determine the final outcome. Following the implementation of the GSVR+PEPNet, we observed a \emph{\textbf{4.6\%}} increase in cumulative \emph{\textbf{Revenue}} and a \emph{\textbf{4.9\%}} rise in user \emph{\textbf{Click-Through Rate (CTR)}} compared to the base model. The online results further corroborate the effectiveness of GSVR.  \\
\textbf{Parameters Cost.} Assume that the original representation parameter memory is M. Not every user and item appears in every scenario, the newly added scenario-specific representation parameters are 1.2M in four scenarios. We quantified the scenario-specific representations and compressed the parameter memory to 0.28M.

\section{Conclusions}
In this paper, we introduce the Global-Distribution Aware Scenario-Specific Variational Representation Learning Framework (GSVR), designed to be directly applicable to existing multi-scenario methods. Specifically, GSVR employs a probabilistic model to generate scenario-specific distributions for each user and item in each scenario. Furthermore, we introduce the global-knowledge aware multinomial-distributions as prior knowledge to regulate the model learning, ensuring the resemblance between the distributions of users with similar interests and items with comparable auxiliary information. Ultimately, experimental results validate the effectiveness of GSVR in enhancing the ability of modeling multi-scenarios.


\begin{thebibliography}{99}

\bibitem{tensorflow}Martín Abadi, Paul Barham, Jianmin Chen, Zhifeng Chen, Andy Davis, Jeffrey Dean, Matthieu Devin, Sanjay Ghemawat, Geoffrey Irving, Michael Isard, et al. 2016. Tensorflow: A system for large-scale machine learning. In \emph{12th {USENIX} symposium on operating systems design and implementation ({OSDI} 16)}.

\bibitem{cross}Alex Beutel, Paul Covington, Sagar Jain, Can Xu, Jia Li, Vince Gatto, and Ed H. Chi. 2018. Latent Cross: Making Use of Context in Recurrent Recommender Systems. In \emph{Proceedings of the Eleventh ACM International Conference on Web Search and Data Mining (WSDM)}. (Feb. 2018), 46-54.

\bibitem{mean}David M Blei, Alp Kucukelbir, and Jon D McAuliffe. 2017. Variational inference: A review for statisticians. \emph{Journal of the American statistical Association}. 112, 518 (2017), 859–877.

\bibitem{causalint}Yichao Wang, Huifeng Guo , Bo Chen, Weiwen Liu, Zhirong Liu, Qi Zhang, Zhicheng He, Hongkun Zhen, Weiwei Yao, Muyu Zhang, Zhenhua Dong, and Ruiming Tang. 2022. CausalInt: Causal Inspired Intervention for Multi-Scenario Recommendation. In \emph{Proceedings of the 28th ACM SIGKDD Conference on Knowledge Discovery and Data Mining (KDD)}. (Aug. 2022), 4090-4099.

\bibitem{pepnet}Jianxin Chang, Chenbin Zhang, Yiqun Hui, Dewei Leng, Yanan Niu, Yang Song, and Kun Gai. 2023. PEPNet: Parameter and Embedding Personalized Network for Infusing with Personalized Prior Information. In \emph{Proceedings of the 29th ACM SIGKDD Conference on Knowledge Discovery and Data Mining (KDD)}. (Aug. 2023), 3795-3804.

\bibitem{deepfm}Huifeng Guo, Ruiming Tang, Yunming Ye, Zhenguo Li, and Xiuqiang He. 2017. Deepfm: a factorization-machine based neural network for ctr prediction. In \emph{Proceedings of the 26th International Joint Conference on Artificial Intelligence (IJCAI)}. Melbourne, Australia., 2782–2788.

\bibitem{dffm}Wei Guo, Chenxu Zhu, Fan Yan, Bo Chen, Weiwen Liu, Huifeng Guo, Hongkun Zheng, Yong Liu, and Ruiming Tang. 2023. DFFM: Domain Facilitated Feature Modeling for CTR Prediction. In \emph{Proceedings of the Proceedings of the 32nd ACM International Conference on Information and Knowledge Management (CIKM)}. (Oct. 2023), 4602-4608.

\bibitem{xavier}Xavier Glorot and Yoshua Bengio. 2010. Understanding the difficulty of training deep feedforward neural networks. In \emph{Proceedings of the thirteenth international conference on artificial intelligence and statistics}. 249–256.

\bibitem{res}Kaiming He, Xiangyu Zhang, Shaoqing Ren, and Jian Sun. 2016. Deep Residual Learning for Image Recognition. In \emph{2016 IEEE Conference on Computer Vision and Pattern Recognition (CVPR)}. (Jun. 2016), 770-778.

\bibitem{b_vae}Irina Higgins, Lo¨ıc Matthey, Arka Pal, Christopher Burgess, Xavier Glorot, Matthew Botvinick, Shakir Mohamed, and Alexander Lerchner. 2017. beta-VAE: Learning Basic Visual Concepts with a Constrained Variational Framework. In \emph{Proceedings of the 5th International Conference on Learning Representations (ICLR)}. (Apr. 2017).

\bibitem{samd}Zhaoxin Huan, Ang Li, Xiaolu Zhang, Xu Min, Jieyu Yang, Yong He, and Jun Zhou. 2023. SAMD: An Industrial Framework for Heterogeneous Multi- Scenario Recommendation. In \emph{Proceedings of the 29th ACM SIGKDD Con- ference on Knowledge Discovery and Data Mining (KDD)}. (Aug. 2023),  4175-4184.

\bibitem{mtl2} Alex Kendall, Yarin Gal, and Roberto Cipolla. 2018. Multi-task learning using uncertainty to weight losses for scenario geometry and semantics. In \emph{Proceedings of the IEEE Conference on Computer Vision and Pattern Recognition}. UT, USA, 7482–7491.

\bibitem{adam} Diederik P. Kingma and Jimmy Ba. 2015. Adam: A Method for Stochastic Optimization. In ICLR.

\bibitem{vae}Diederik P. Kingma, Max Welling. 2014. Auto-Encoding Variational Bayes. In \emph{Proceedings of the 2nd International Conference on Learning Representations (ICLR)}. (Apr. 2014).

\bibitem{adl} Jinyun Li, Huiwen Zheng, Yuanlin Liu, Minfang Lu, Lixia Wu, and Haoyuan Hu. 2023. ADL: Adaptive Distribution Learning Framework for Multi-Scenario CTR Prediction. In \emph{Proceedings of the 46th International ACM SIGIR Conference on Research and Development in Information Retrieval (SIGIR)}. (Jul. 2023), 1786-1790.

\bibitem{hmoe} Pengcheng Li, Runze Li, Qing Da, An-Xiang Zeng, and Lijun Zhang. 2020. Improving Multi-Scenario Learning to Rank in E-commerce by Exploiting Task Relationships in the Label Space. In \emph{Proceedings of the 29th ACM International Conference on Information and Knowledge Management (CIKM)}. (Oct. 2020), 2605-2612.

\bibitem{mtl1}Pengfei Liu, Xipeng Qiu, and Xuanjing Huang. 2017. Adversarial Multi-task Learning for Text Classification. In \emph{Proceedings of the 55th Annual Meeting of the Association for Computational Linguistics}. Vancouver, Canada, 1–10.

\bibitem{mtl3}Jiaqi Ma, Zhe Zhao, Xinyang Yi, Jilin Chen, Lichan Hong, and Ed H Chi. 2018. Modeling task relationships in multi-task learning with multi-gate mixture-of-experts. In \emph{Proceedings of the 24th ACM SIGKDD International Conference on Knowledge Discovery \& Data Mining (KDD)}. London, UK, 1930–1939.

\bibitem{m-test}Simon J Mason and Nicholas E Graham. 2002. Areas beneath the relative operating characteristics (ROC) and relative operating levels (ROL) curves: Statistical significance and interpretation. \emph{Quarterly Journal of the Royal Meteorological Society: A journal of the atmospheric sciences, applied meteorology and physical oceanography}. 128, 584 (2002), 2145–2166.

\bibitem{hc2}Shanlei Mu, Penghui Wei, Wayne Xin Zhao, Shaoguo Liu, Liang Wang, Bo Zheng. 2023. Hybrid Contrastive Constraints for Multi-Scenario Ad Ranking. In \emph{Proceedings of the 32nd ACM International Conference on Information and Knowledge Management (CIKM)}. (Oct. 2023), 1857-1866.

\bibitem{back1} Badrul Sarwar, George Karypis, Joseph Konstan, and John Riedl. 2001. Item-based collaborative filtering recommendation algorithms. In \emph{Proceedings of the 10th international conference on World Wide Web}. 285–295.

\bibitem{moe} Noam Shazeer, Azalia Mirhoseini, Krzysztof Maziarz, Andy Davis, Quoc Le, Geoffrey Hinton, and Jeff Dean. 2017. Outrageously Large Neural Networks: The Sparsely-Gated Mixture-of-Experts Layer. In \emph{Proceedings of the 5th International Conference on Learning Representations (ICLR)}. (Apr. 2017).

\bibitem{sar} Qijie Shen, Wanjie Tao, Jing Zhang, Hong Wen, Zulong Chen, and Quan Lu. 2021. SAR-Net: A Scenario-Aware Ranking Network for Personalized Fair Recommendation in Hundreds of Travel Scenarios. In \emph{Proceedings of the 30th ACM International Conference on Information and Knowledge Management (CIKM)}. (Nov. 2021), 4094-4103.

\bibitem{lhuc} Pawel Swietojanski, Jinyu Li, and Steve Renals. 2016. Learning hidden unit contributions for unsupervised acoustic model adaptation. \emph{IEEE/ACM Transactions on Audio, Speech, and Language Processing}. 24, 8 (2016), 1450–1463.

\bibitem{star}Xiang-Rong Sheng, Liqin Zhao, Guorui Zhou, Xinyao Ding, Binding Dai, Qiang Luo, Siran Yang, Jingshan Lv, Chi Zhang, Hongbo Deng, et al. 2021. One model to serve all: Star topology adaptive recommender for multi-domain ctr prediction. In \emph{Proceedings of the 30th ACM International Conference on Information \& Knowledge Management (CIKM)}. 4104–4113.

\bibitem{ple} Hongyan Tang, Junning Liu, Ming Zhao, and Xudong Gong. 2020. Progressive layered extraction (ple): A novel multi-task learning (mtl) model for personalized recommendations. In \emph{Fourteenth ACM Conference on Recommender Systems}. 269– 278.

\bibitem{maria} Yu Tian, Bofang Li, Si Chen, Xubin Li, Hongbo Deng, Jian Xu, Bo Zheng, Qian Wang, and Chenliang Li. 2023. Multi-Scenario Ranking with Adaptive Feature Learning. In \emph{Proceedings of the 46th International ACM SIGIR Conference on Research and Development in Information Retrieval (SIGIR)}. (Jul. 2023), 517-526.

\bibitem{trans} Ashish Vaswani, Noam Shazeer, Niki Parmar, Jakob Uszkoreit, Llion Jones, Aidan N Gomez, Łukasz Kaiser and Illia Polosukhin. 2017. Attention is All you Need. In \emph{Proceedings of 30th Conference on Advances in Neural Information Processing Systems (NIPS)}. (Dec. 2017), 5998–6008.

\bibitem{back2} Hong Wen, Jing Zhang, Fuyu Lv, Wentian Bao, Tianyi Wang, and Zulong Chen. 2021. Hierarchically Modeling Micro and Macro Behaviors via Multi-Task Learn- ing for Conversion Rate Prediction. In \emph{Proceedings of the 44th International ACM SIGIR Conference on Research and Development in Information Retrieval (SIGIR)}.

\bibitem{back3} Hong Wen, Jing Zhang, Yuan Wang, Fuyu Lv, Wentian Bao, Quan Lin, and Keping Yang. 2020. Entire space multi-task modeling via post-click behavior decom- position for conversion rate prediction. In \emph{Proceedings of the 43rd International ACM SIGIR Conference on Research and Development in Information Retrieval (SIGIR)}. 2377–2386.

\bibitem{plate} Yuhao Wang, Xiangyu Zhao, Bo Chen, Qidong Liu, Huifeng Guo, Huan-shuo Liu, Yichao Wang, Rui Zhang, and Ruiming Tang. 2023. PLATE: A Prompt-Enhanced Paradigm for Multi-Scenario Recommendations. In \emph{Proceedings of the 46th International ACM SIGIR Conference on Research and Development in Information Retrieval (SIGIR)}. (Jul. 2023), 1498-1507.

\bibitem{adasparse} Xuanhua Yang, Xiaoyu Peng, Penghui Wei, Shaoguo Liu, Liang Wang and Bo Zheng. 2022. AdaSparse: Learning Adaptively Sparse Structures for Multi-Domain Click-Through Rate Prediction. In \emph{Proceedings of the 31st ACM International Conference on Information and Knowledge Management (CIKM)}. (Oct. 2022), 4635-4639.

\bibitem{back4} Jing Zhang and Dacheng Tao. 2021. Empowering Things With Intelligence: A Survey of the Progress, Challenges, and Opportunities in Artificial Intelligence of Things. \emph{IEEE Internet of Things Journal}. 8(10), 7789–7817. 

\bibitem{m2m} Qianqian Zhang, Xinru Liao, Quan Liu, Jian Xu, Bo Zheng. 2022. Leaving No One Behind: A Multi-Scenario Multi-Task Meta Learning Approach for Advertiser Modeling. In \emph{Proceedings of the Fifteenth ACM International Conference on Web Search and Data Mining (WSDM)}. (Feb. 2022), 1368-1376.

\bibitem{m5}Pengyu Zhao, Xin Gao, Chunxu Xu, and Liang Chen. 2023. M5: Multi-Modal Multi-Interest Multi-Scenario Matching for Over-the-Top Recommendation. In \emph{Proceedings of the 29th ACM SIGKDD Conference on Knowledge Discovery and Data Mining (KDD)}. (Aug. 2023), 5650-5659.

\bibitem{3mn} Yifei Zhang, Hua Hua, Hui Guo, Shuangyang Wang, Chongyu Zhong, and Shijie Zhang. 2023. 3MN: Three Meta Networks for Multi-Scenario and Multi-Task Learning in Online Advertising Recommender Systems. In \emph{Proceedings of the 32nd ACM International Conference on Information and Knowledge Management (CIKM)}. (Oct. 2023), 4945-4951.

\bibitem{sass} Yuanliang Zhang, Xiaofeng Wang, Jinxin Hu, Ke Gao, Chenyi Lei, and Fei Fang. 2022. Scenario-Adaptive and Self-Supervised Model for Multi-Scenario Personalized Recommendation. In \emph{Proceedings of the 31st ACM International Conference on Information and Knowledge Management (CIKM)}. (Oct. 2022), 3674-3683.

\bibitem{sfpnet} Moyu Zhang, Yongxiang Tang, Jinxin Hu, and Yu Zhang. 2024. Scenario-Adaptive Fine-Grained Personalization Network: Tailoring User Behavior Representation to the Scenario Context. In \emph{Proceedings of the 47th International ACM SIGIR Conference on Research and Development in Information Retrieval (SIGIR)}. (Jul. 2024).

\bibitem{din} Guorui Zhou, Xiaoqiang Zhu, Chenru Song, Ying Fan, Han Zhu, Xiao Ma, Yanghui Yan, Junqi Jin, Han Li, and Kun Gai. 2018. Deep interest network for click-through rate prediction. In \emph{Proceedings of the 24th ACM SIGKDD International Conference on Knowledge Discovery \& Data Mining (KDD)}. 1059–1068.

\bibitem{hinet} Jie Zhou, Xianshuai Cao, Wenhao Li, Lin Bo, Kun Zhang, Chuan Luo, and Qian Yu. 2023. HiNet: Novel Multi-Scenario \& Multi-Task Learning with Hierarchical Information Extraction. In  \emph{Proceedings of the 39th IEEE International Conference on Data Engineering (ICDE)}. (Apr. 2023), 2969-2975. 

\bibitem{aesm2} Xinyu Zou, Zhi Hu, Yiming Zhao, Xuchu Ding, Zhongyi Liu, Chenliang Li, Aixin Sun. 2022. Automatic Expert Selection for Multi-Scenario and Multi- Task Search. In \emph{Proceedings of the 45th International ACM SIGIR Conference on Research and Development in Information Retrieval (SIGIR)}. (Jul. 2022), 1535-1544.



\end{thebibliography}
\end{document}